\documentclass[twocolumn]{aastex631}

\graphicspath{{./}{figures/}}
\usepackage{graphicx}	
\usepackage{amsmath}	
\usepackage{amssymb}	
\usepackage{float,hyperref}
\usepackage{color,ulem}
\usepackage{natbib}
\newcommand{\K}{\, \rm K}

\newcommand{\km}{\, \rm km}
\newcommand{\m}{\, \rm m}
\newcommand{\s}{\, \rm s}
\newcommand{\AU}{\, \rm AU}
\newcommand{\pc}{\, \rm pc}
\newcommand{\mas}{\, \rm mas}
\newcommand{\cm}{\, \rm cm}
\newcommand{\g}{\, \rm g}

\newcommand{\CO}{$^{12}$CO\,}

\newcommand{\incA}{41}
\newcommand{\PAA}{40}
\newcommand{\RcA}{3.7}

\newcommand{\incAerr}{$^{+13}_{-7}$}
\newcommand{\PAAerr}{$^{+20}_{-10}$}
\newcommand{\RcAerr}{$^{+0.3}_{-0.2}$}

\newcommand{\incB}{46}
\newcommand{\PAB}{40}
\newcommand{\RcB}{3.6}

\newcommand{\incBerr}{$\pm 9$}
\newcommand{\PABerr}{$^{+8}_{-14}$}
\newcommand{\RcBerr}{$^{+0.8}_{-0.6}$}

\definecolor{lightgreen}{rgb}{0, 0.7, .6}

\definecolor{slate}{rgb}{0.35, 0.48, 0.59}

\newcommand{\x}{\sout}

\begin{document}
\title{Sites of Planet Formation in Binary Systems. II. Double the Disks in DF Tau}

\author[0000-0002-7219-0064]{Taylor Kutra}
\affiliation{Lowell Observatory, Flagstaff, AZ 86001 USA}

\author[0000-0001-7998-226X]{Lisa Prato}
\affiliation{Lowell Observatory, Flagstaff, AZ 86001 USA}

\author[0000-0003-2053-0749]{Benjamin M Tofflemire}
\altaffiliation{51 Pegasi b Fellow}
\affiliation{Department of Astronomy, The University of Texas at Austin, Austin, TX 78712, USA}

\author[0000-0001-9674-1564]{Rachel Akeson}\affiliation{IPAC/Caltech, Pasadena, CA, 91125, USA}

\author[0000-0001-5415-9189]{G. H. Schaefer}
\affiliation{The CHARA Array of Georgia State University, Mount Wilson Observatory, Mount Wilson, CA 91023, USA}

\author[0000-0003-4247-1401]{Shih-Yun Tang} 
\affiliation{Physics \& Astronomy Department, Rice University, 6100 Main St., Houston, TX 77005, USA}
\affiliation{Lowell Observatory, Flagstaff, AZ 86001 USA}

\author[0000-0003-3172-6763]{Dominique Segura-Cox} \affiliation{Department of Astronomy, The University of Texas at Austin, Austin, TX 78712, USA}

\author[0000-0002-8828-6386]{Christopher M. Johns-Krull}
\affiliation{Physics \& Astronomy Department, Rice University, 6100 Main St., Houston, TX 77005, USA}

\author[0000-0001-9811-568X]{Adam Kraus}
\affiliation{Department of Astronomy, The University of Texas at Austin, Austin, TX 78712, USA}

\author[0000-0003-2253-2270]{Sean Andrews}
\affiliation{Center for Astrophysics, Harvard \& Smithsonian, 60 Garden Street, Cambridge, MA 02138, USA}

\author[0000-0002-4625-7333]{Eric L. N. Jensen}\affiliation{Dept. of Physics \& Astronomy, Swarthmore College, 500 College Ave., Swarthmore, PA 19081, USA}

\shorttitle{Double the Disks in DF Tau}
\shortauthors{Kutra et al.}

\correspondingauthor{Taylor Kutra}
\email{tkutra@lowell.edu}

\begin{abstract}
    This article presents the latest results of our ALMA program to study circumstellar disk characteristics as a function of orbital and stellar properties in a sample of young binary star systems known to host at least one disk. Optical and infrared observations of the eccentric, $\sim$48-year period binary DF Tau indicated the presence of only one disk around the brighter component. However, our $1.3 {\rm mm}$ ALMA thermal continuum maps show two nearly-equal brightness components in this system. We present these observations within the context of updated stellar and orbital properties which indicate that the inner disk of the secondary is absent. Because the two stars likely formed together, with the same composition, in the same environment, and at the same time, we expect their disks to be co-eval. However the absence of an inner disk around the secondary suggests uneven dissipation. We consider several processes which have the potential to accelerate inner disk evolution. Rapid inner disk dissipation has important implications for planet formation, particularly in the terrestrial-planet-forming region. 
\end{abstract}


\section{Introduction} \label{sec:intro}

A comprehensive paradigm for planet formation requires a fundamental understanding of the conditions, process, and timescale for circumstellar disk evolution. 
Although advancements in theory \citep[see review by][]{Pascucci2023} and observations, such as ALMA \citep{Barenfeld2016,Gudel2018} or JWST \citep{Bajaj2024}, have brought about new insights, puzzling challenges remain. 
For example, what triggers the onset of disk dissipation, what is the source of the $\alpha$ viscosity in the disk, and what sustains the persistence of some primordial disks for 5--10 Myr \citep{Herczeg2023}?

Young binary systems offer a unique opportunity to study both the fragility and robustness of circumstellar disks because membership in a binary system dictates that tidal truncation by the companion \citep[e.g.,][]{AL1994} limits the disk's outer radii. Thus disks in binaries have determinate maximum sizes.
Given the dynamical complexity, it is unsurprising that circumstellar disks in binaries with orbital separations of $\lesssim 100 \AU$ are smaller, less massive \citep{Jensen1996,Harris2012,Barenfeld2019}, and shorter-lived \citep{Cieza2009,Kraus2012,Cheetham2015,Barenfeld2019} than disks around more widely spaced or single stars \citep[e.g.,][]{Akeson2019}. 

Yet despite these seemingly bleak prospects for disk longevity and potential planet formation, in some cases the circumstellar disks in close binary systems persist. Furthermore, planet formation can proceed in close binary systems, as exemplified by $\gamma$ Cephei b \citep{Hatzes2003}, GJ 86b \citep{Queloz2000} and HD 196885 Ab \citep{Correia2008}, albeit less frequently \citep{Kraus2016}. Clearly circumstellar disk evolution is not regulated entirely by membership in a close binary; other processes and influences are at work. 
This is demonstrated by the large range in age at which circumstellar disks dissipate \citep[e.g.,][]{Haisch2001,Williams2011}. 
In addition to the binary environment, other factors drive disk dissipation: by studying the stellar, orbital, and disk properties in young binaries, we can use these nominally coeval systems to explore these factors, interpreting the evolution of all primordial disks through this lens.

Using this paradigm, we investigate the young (1-2 Myr) visual binary, DF Tau. It is composed of two roughly equal mass M2 stars on a 48 year orbit in the Taurus star forming region \citep{Allen2017}. The projected orbital separation is only $\sim100 \mas$, corresponding to a physical separation of $\sim14 \AU$ at a distance of $142 \pc$ \citep{Krolikowski2021}. Both the A and B components were designated as classical T Tauri stars (CTTSs) by \citet{Hartigan2003} and \citet{White2001} based on accretion signatures (i.e., the H$\alpha$ equivalent width, UV excess, [O I] molecular emission, and veiling at $6100 \dot{A}$). 

\citet{White2001} found a much weaker ultraviolet (UV) excess for the DF Tau secondary, $\Delta U = 0.85$ mag, compared to that of the primary, $\Delta U = 2.43$ mag.  Their criterion for the presence of a disk is $\Delta U = 0.8$ mag; however, the uncertainty in the U-band magnitude is $\sim 30\%$ (0.29 mag), indicating a marginal UV CTTS assignation for DF Tau B. 

\citet{Allen2017} presented component-resolved, near-infrared (NIR) spectroscopy and photometry of the DF Tau binary and concluded that for DF Tau A, the NIR colors, IR veiling, and optical/NIR colors were all consistent with the presence of an accreting circumstellar disk. However, they found little variability and no indication of accretion or significant amounts of circumstellar material around the secondary and cautioned that the disk signatures seen for DF Tau B in \citet{Hartigan2003} could have been the result of contamination from the disk of DF Tau A, given that the angular separation at the time of observation was close to the resolution limit of the \citet{Hartigan2003} HST STIS observations.

DF Tau is part of our young, small-separation binary sample observed with ALMA in Cycle 7 (PI Tofflemire). Targets were selected for this program on the basis of their relatively complete orbital solutions. As demonstrated in our recent study of the FO Tau binary \citep{Tofflemire2024}, our goal is to understand the disk dissipation process in the context of the binaries' stellar and orbital properties. This work is also a component of our larger exploration of the spectroscopic and photometric properties of the individual components in a sample of close to 100 young binary systems \citep[e.g.,][]{Kellogg2017,Allen2017,Sullivan2019,Prato2023b}. 

In this contribution we focus on our new ALMA observations, reanalyze the DF Tau NIR component spectra, including measurements of the stars' surface averaged magnetic fields, and update the orbital solution with our most recent adaptive optics (AO) observations. Given the precise orbital solution, stellar parameters that suggest similar component temperatures and masses, and clear evidence for an inner disk around the primary with only marginal evidence for an inner disk surrounding the secondary star, DF Tau is a key target for understanding the onset of disk dissipation. 

In section \ref{sec:obs} we describe our observations and data reduction from ALMA, Keck, TESS/K2, and the Lowell 0.7-m and 1.1m telescopes and we present our analysis of the multi-wavelength dataset in Section \ref{sec:analysis}. 
In Section \ref{sec:discuss} we discuss our results in terms of the disk and binary interaction, the stellar spin and obliquity and possible origins for the differences contributing to disk dissipation in DF Tau. We summarise our work and list main conclusions in Section \ref{sec:summ}.

\section{Observations and Data Reduction}\label{sec:obs}

\subsection{ALMA}\label{subsec:ALMA}
Our Cycle 7 observations of DF Tau (project code 2019.1.01739.S) mirror those of \citet{Tofflemire2024}. We used Band 6 receivers in dual polarization mode and sampled continuum emission over three spectral windows centered on 231.6, 244.0, and 245.9 GHz each with a bandwidth of 1.875 GHz and 31.25 MHz resolution. In order to resolve the Keplerian rotation profile of the disk, we chose a fourth spectral window that covers the \CO $J=2-1$ transition at 230.538 GHz for the source's heliocentric velocity \citep[$16.4 \, ~{\rm km}/{\rm s}$,][]{Kounkel2019} 

Observations were taken in compact (short baselines, SB) and extended (long baselines, LB) array configurations to ensure both high angular resolution and sensitivity to extended emission. The compact configuration ($\sim $C6 configuration), with baselines between 14 m and 3.6 km, achieved an angular resolution and maximum recoverable scale of 0.1" and 2.1", respectively. Observations took place on 18 July 2021 UTC for $332.640\s$ using the compact configuration. The extended configuration ($\sim$ C8 configuration), with baselines between 40m and 11.6 km, achieved an angular resolution of 0.031" and a maximum recoverable scale of 0.67". The observations for this configuration took place on 24 August 2021 UTC for $2830.464\s$ in the extended configuration. The minimum and maximum angular scales covered by both the compact and extended configuration translate to $\sim4.6$ to $100 \AU$ at the distance of Taurus \citep[$\sim 140 \pc$][]{Kenyon1994,Krolikowski2021}. Given this resolution, we were able to separate the emission from the two components and marginally resolve the individual disks. The angular separation between the components is $\sim 0.1$" which corresponds to a physical separation of $14 \AU$ and a tidal truncation radius of $\lesssim 5\AU$ (0.04") for a binary of roughly equal mass in a circular orbit \citep{AL1994}.

\subsubsection{Calibration and Imaging}

\begin{figure*}
    \vspace{20pt}
    \centering
    \includegraphics[width=\textwidth]{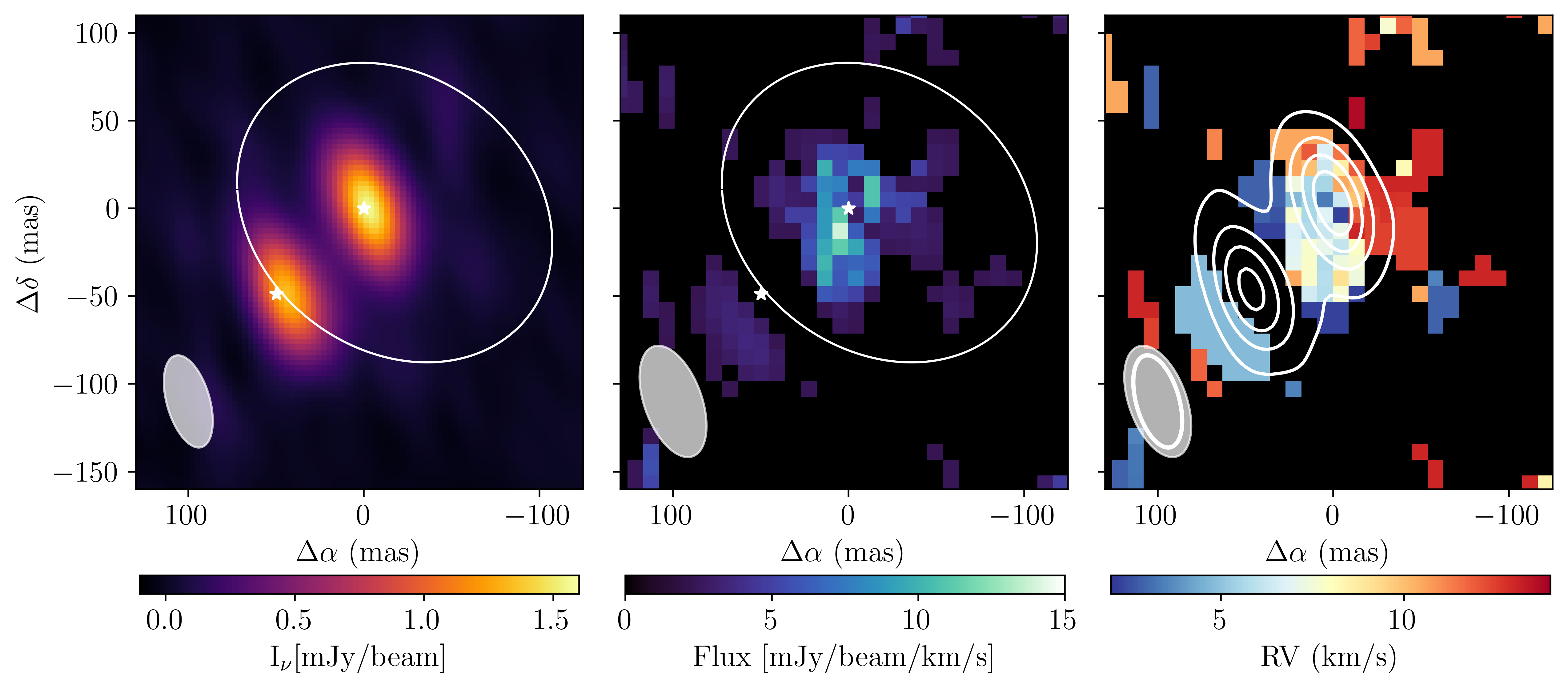}
    \caption{Continuum intensity (left), \CO line integrated intensity (moment-0; middle, RMS of 0.8 mJy/beam) and \CO first moment maps (right) show detections of circumstellar dust and gas orbiting both the primary and secondary stars of DF Tau. In the left and middle panels, the relative binary orbit of the secondary around the primary is overplotted in white and the centers of the continuum emission, determined using \texttt{imfit}, are shown as white stars, confirming alignment with the binary orbit. 
    The first-moment map (right, continuum emission and beam size are shown with overplotted white contours, hatched region in the color bar indicates velocities which are highly obscured by molecular cloud absorption) shows rough agreement in the direction of rotation of the disks. Beam sizes are shown for all panels in bottom left. }
    \label{fig:alma}
\end{figure*}

The extended and compact configurations were both calibrated by the standard ALMA pipeline \citep[\texttt{CASA v6.1.15}, ][]{McMullin2007}. The compact configuration was provided as a fully calibrated measurement set by the North American ALMA Science Center. The extended configuration, however, had several antennae with poor responses to the calibrator in the QA0 report. To ensure these antennae did not introduce any anomalous correlated noise features, we flagged these 4 additional antennae from the long baseline measurement set before rerunning the calibration and imaging pipelines. 

After the raw data was appropriately calibrated, we followed post-processing procedures developed for the DSHARP large ALMA program and focused on combining array configurations and self-calibrating \citep{Andrews2018}. A full description of this approach with accompanying scripts can be found on the program’s data release web page\footnote{\href{https://almascience.eso.org/almadata/lp/DSHARP/}{https://almascience.eso.org/almadata/lp/DSHARP/}}.

We first created a continuum data set for each configuration by excluding channels within $\pm 25 \km/\s$ of the \CO $\,J=2-1$ transition at the DF Tau systematic velocity. We then imaged each data set using \texttt{tclean} (briggs weighting, \texttt{robust}$=0$) to test the astrometric alignment and shift both measurement sets to align the phase centers on the DF Tau A continuum source. Finally, we re-scaled the compact configuration measurement set's flux to match that of the extended measurement set. 

Then we began to iterate on the phase-only self-calibrations for the compact configuration. After using \texttt{tclean} without a mask to create an initial image, we began the self-calibrations with a mask which encompasses only the largest flux in the dirty map so as to not give legitimacy to any spurious correlated noise signals. This created a source model which we used to calibrate the visibilities. We then used two large circular apertures centered on the sources to calculate the signal-to-noise ratio (SNR) for a measure of the strength of the signal before self-calibration. We successively iterated on the mask for each round of self-calibration, ensuring that only regions with high fluxes were masked. 

We continued to image, adjust the mask, and self-calibrate until either the SNR did not improve from the previous iteration, the number of flagged solutions for any time interval exceeded $20\%$ (to avoid models which insufficiently describe the data), or we reached the native integration step ($6 s$). For the compact configuration, solution intervals of the total duration and of $8000 s$ increased the SNR from an initial value of $41$ to a pleateau of $\sim 73$. These solution intervals use the entire integration time to compute the gain. Amplitude-only self calibration did not improve the SNR and therefore we continued without it. 

We then combined the self-calibrated compact measurement set with the data from the extended configuration and repeated the same process as above. Establishing the initial mask for the combined measurement set after the initial cleaning was less straightforward than for the compact measurement set because the correlated noise in the beam falls along the same position angle as the anticipated location of the secondary star's disk. To avoid including any spurious signal in our masks, we experimented with masking only the signal at the expected position of the primary and then masking the signal at the expected positions of both the primary and secondary during the first round of phase-only self calibration. The correlated noise features were significantly reduced with masks on the primary and secondary. Therefore we continued the phase-only self-calibrations using an initial mask which encompassed only the areas with the highest flux and centered on the predicted locations of the primary and secondary stars. Applying self-calibration for the full duration and for $8000\s$ increased the SNR from $29$ to $117$. Shorter intervals and amplitude-only self calibration did not significantly improve the data quality, therefore we stopped iterating. 

The final continuum imaging, Figure \ref{fig:alma}, was performed with fully interactive masking using Briggs weighting (robust = 0) with a three sigma cleaning threshold. The synthesized continuum beam was $0.054" \times 0.025 " $ with a position angle of $16.02^\circ$. The root-mean-square deviation (RMS) of the final image was $0.013$ mJy$/$beam. 

We also created $^{12}$CO $J=2-1$ data cubes, using channels within $\pm 15 \km/\s$ of DF Tau's systemic velocity, for the compact and extended configurations from the initial measurement sets. We then applied the self-calibration steps described above and subtracted the continuum emission. We used a 2-channel spacing of $0.635 \km/\s$. The final cube was imaged with interactive masking, using Briggs robust$=1$ weighting. The synthesized CO beam is $ 0.066" \times 0.033" $ mas with a position angle of $19.84 ^{\circ}$. The CO detection is marginal and the RMS of the channel maps is $\sim1$ mJy/beam. Therefore we only present the integrated intensity and first moment maps in Figure \ref{fig:alma}. The RMS for the integrated intensity map is $0.8$ mJy/beam. Cloud absorption is apparent in both moment maps, particularly for the north-west (red-shifted) side of the secondary disk. The velocity channels which are most affected have LSRK radio velocities between 5-10 km/s (denoted in Figure \ref{fig:alma} with hatched region in the right-most panel’s velocity color bar). The LSRK velocity of the Taurus molecular cloud \citep[$\sim$ 5-8 km/s][]{Narayanan2008} lies within this same range. 

\subsection{Keck -- AO Imaging}\label{subsec:keck-nirc2}
We obtained spatially resolved images of DF Tau on three nights between 2019 through 2022 using the NIRC2 camera and the adaptive optics (AO) system on the Keck II Telescope \citep{Wizinowich2000}. On each night we obtained 9$-$12 images, dithered by $2\farcs$, in each of the narrow-band Hcont (central wavelength 1.5804 $\mu$m) and Kcont (2.2706 $\mu$m) filters. Additional images were obtained using the Jcont (1.2132 $\mu$m) and Lp (3.776 $\mu$m) filters on UT 2022 December 29, however, the Lp images were saturated. Each image consisted of 10 coadded exposures of 0.2$-$0.8 sec. We flat fielded the images using dark-subtracted dome flats and removed the sky background by subtracting pairs of dithered images. 

We used the single star DN Tau as a point-spread function reference (PSF) observed with the same AO frame-rate immediately before or after the observations of DF Tau. DN Tau has been used as a PSF reference in earlier AO studies \citep{Schaefer2006} and no companion was detected in prior multiplicity surveys \citep{Simon1995,Kraus2011}. We constructed binary models using the PSF grid search procedure described by \citet{Schaefer2014} to measure the separation ($\rho$), position angle (P.A.) east of north, and flux ratio ($f_B/f_A$) of the components in DF Tau. We applied the geometric distortion solution computed by \citet{Service2016} and used a plate scale of 9.971 $\pm$ 0.004 mas~pixel$^{-1}$ and subtracted $0\fdg262 \pm 0\fdg020$ from the measured position angles to correct for the orientation of the camera relative to true north. The binary positions and flux ratios are presented in Table~\ref{tab.sepPA}. 

\subsection{Keck Spectroscopy: NIRSPEC Behind AO}\label{subsec:keck}

The spectra presented here were also published in \citet{Allen2017} but we include a description of the observations here for completeness and because some steps in the processing were updated (e.g., continuum normalization).
We used the cross-dispersed, cryogenic, NIR spectrograph NIRSPEC \citep{McLean1998}, deployed behind the Keck II AO system (as NIRSPAO), to obtain angularly-resolved spectra of DF Tau on UT 2009 December 6. In high-spectral resolution mode, we achieved R$=$30,000 using the two-pixel slit (reimaged by the AO system to a width of $0\farcs027\times2\farcs26$ arcseconds) oriented to the position angle of the binary at the epoch of our observations, $\sim$224$^{\circ}$. Natural seeing at the time of observation was better than 0$\farcs6$; The AO frame rate was $\sim$438 s$^{-1}$ with about 500 Wave Front Sensor counts. Four 300 s exposures were taken in an ABBA pattern with a $\sim$1$''$ nod; the airmass was 1.03. The central order 49 of the N5 filter, with echelle and cross disperser settings of 63.04 and 36.3, respectively, yielded a wavelength range of 1.5448--1.5675 $\mu$m. This order is particularly fortuitous as no significant telluric absorption is present at the high altitude, dry site on Mauna Kea, and the combination of atomic and molecular lines provide for the characterization of a range of spectral types and other stellar properties \citet{Tang2024}. Seven other full orders are obtained at our H-band setting; three of these are saturated by the atmosphere and the other four (orders 46, 47, 48, and 50) contain numerous telluric lines.

All data reduction was accomplished with the REDSPEC
package\footnote{https://www2.keck.hawaii.edu/inst/nirspec/redspec.html}, which executes both a spatial and spectral rectification \citep{Kim2015}. We differenced the pairs of A-B exposures and divided by a dark-subtracted flat field. A byproduct of this software is a two-dimensional, rectified, wavelength-calibrated spectrum that includes the spectral traces of both binary components. We used custom code to fit a native point spread function to each stellar trace across the detector for order 49 and extract the individual component spectra. The individual PSFs used for the fit to the binary had a FWHM of $\sim$0.05$''$. The spectra were then flattened, normalized, and corrected for barycentric motion. These 2009 DF Tau A and B NIRSPAO spectra were previously published in \citet{Allen2017} and \citet{Prato2023b}; here we reanalyzed them after performing an improved continuum fit using division by a 5th order polynomial and a 1\% renormalization. Additional details regarding the observations and extraction of our NIRSPAO spectra are available in \citet{Kellogg2017}.

\begin{deluxetable*}{lllllll}
\tabletypesize{\scriptsize}
\tablewidth{0pt}
\tablecaption{Keck NIRC2 Adaptive Optics Measurements of DF Tau  \label{tab.sepPA}}
\tablehead{
\colhead{UT Date} &\colhead{UT Time} & \colhead{Julian Year} & \colhead{$\rho$ (mas)} & \colhead{P.A. ($\degr$)} & \colhead{Filter} & \colhead{Flux Ratio}}
\startdata 
2019Jan20 & 06:40  & 2019.0521  &    75.27  $\pm$   0.59  &   164.05  $\pm$   0.45  & Hcont & 0.809 $\pm$ 0.035 \\
          &        &            &                         &                         & Kcont & 0.517 $\pm$ 0.021 \\
2022Oct19 & 11:46  & 2022.7981  &    65.39  $\pm$   1.14  &   119.53  $\pm$   1.00  & Hcont & 0.700 $\pm$ 0.020 \\
          &        &            &                         &                         & Kcont & 0.484 $\pm$ 0.022 \\
2022Dec29 & 07:09  & 2022.9919  &    65.73  $\pm$   0.90  &   116.68  $\pm$   0.78  & Jcont & 0.854 $\pm$ 0.069 \\
          &        &            &                         &                         & Hcont & 0.762 $\pm$ 0.039 \\
          &        &            &                         &                         & Kcont & 0.507 $\pm$ 0.019 \\
\enddata 
\end{deluxetable*} 

\subsection{Time-series Photometry}\label{subsec:phot}

\begin{figure}
    \centering
    \includegraphics[width=0.45\textwidth]{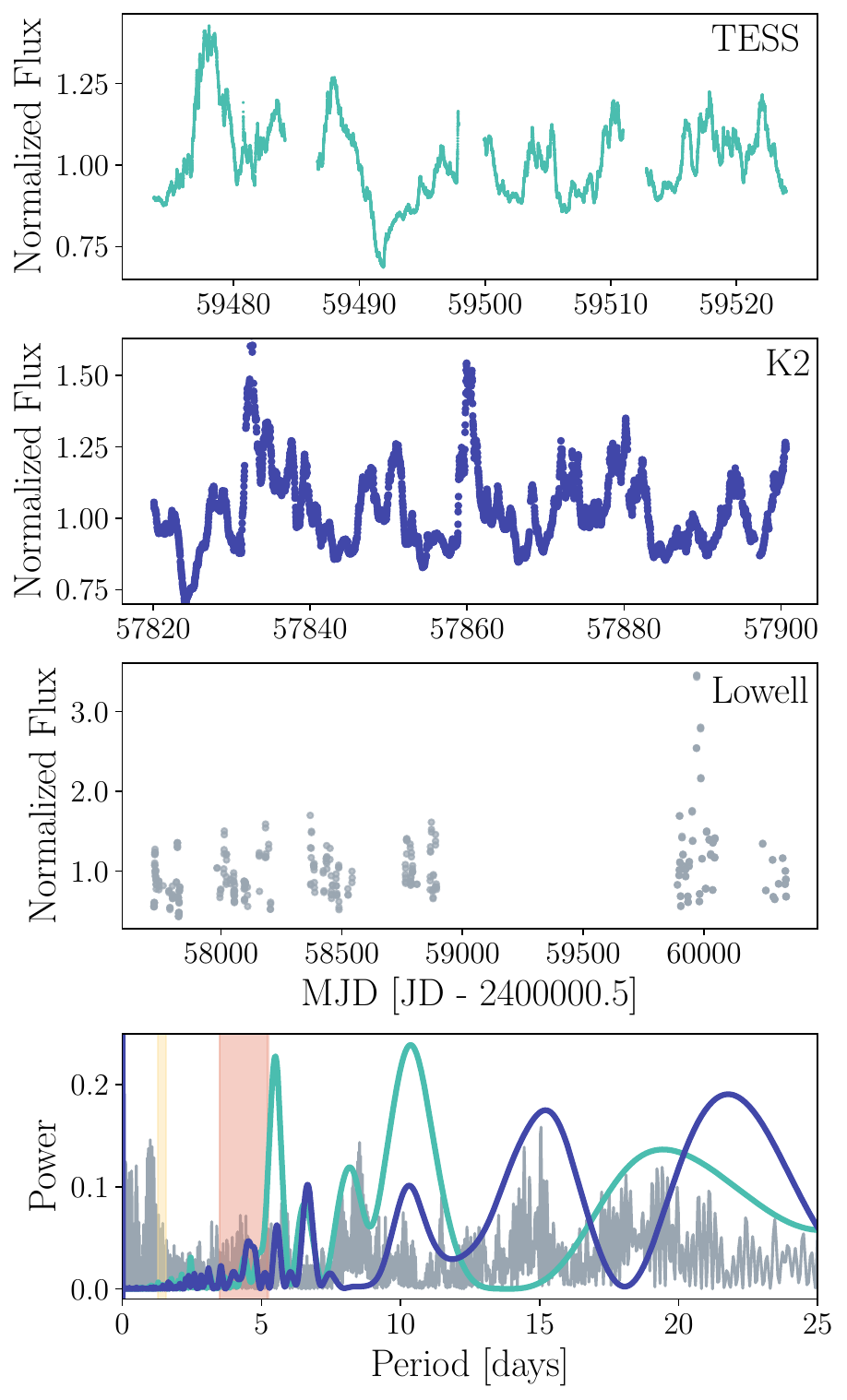}
    \caption{Lightcurves from TESS, K2 and the Lowell 0.7/1.1 m telescopes (top three panels, respectively) and resulting periodograms (bottom panel) for the blended DF Tau system. Maximum rotation periods for the observed $v\sin i$, assuming alignment between stellar obliquity and disk inclination, and theoretical predictions of stellar radii \citep[the 2 Myr tracks from ][]{Feiden2016}, are indicated by shaded regions in red and yellow for the primary and secondary, respectively. There is no significant power at these periods and therefore the variability in the light curve is dominated by stochastic accretion and/or other processes in the inner disk of the primary.}
    \label{fig:periodograms}
\end{figure}

DF Tau was observed by both the Transiting Exoplanet Survey Satellite \citep[TESS; ][]{tess1,tess2} during sectors 43 and 44 (from 16 September through 6 November, 2021 UTC) and $K2$ \citep{k2a,k2b} during Campaign 13 (between 8 March and 27 May, 2017 UTC).
We obtained unresolved time-series photometry of DF Tau from these missions using \texttt{lightkurve} \citep{lightkurve} from the Mikulski Archive for Space Telescopes\footnote{\href{https://archive.stsci.edu}{https://archive.stsci.edu}}. From the TESS mission, we acquired the $120\s$ cadence light curve processed by the Science Processing Operations Center \citep[SPOC][]{spoc}. From $K2$, we obtained the $1800\s$ cadence photometry, which was extracted using the technique described in \citet{Vanderburg2014} that accounts for the systematic errors and mechanical failures of the \textit{Kepler} spacecraft. These lightcurves, and their corresponding Lomb-Scargle periodograms \citep{Lomb1976,Scargle1982}, computed using astropy \citep{astropy:2013,astropy:2018,astropy:2022}, are presented in Figure \ref{fig:periodograms}.

Ground-based photometry of DF Tau continued following the publication of results from an initial season in \citet{Allen2017}. We continued to use the Lowell robotic 0.7-m telescope + CCD and V-band filter through the 2019-2020 observing season. Following closure of that telescope, monitoring was resumed
in 2022 Nov using the Lowell Hall 1.1-m telescope + CCD and V-band filter. The
data were reduced in the same way as in \citet{Allen2017} using conventional differential aperture photometry and the same three comparison stars throughout: HD 283654 (K2III), GSC 1820-0482 (reddened F5/8V), and GSC 1820-0950 (reddened K0:III). While the 0.7-m data usually involved multiple visits each night, the 1.1-m data generally were taken during a single nightly visit that included three to five images. Nightly per-observation uncertainties were $\sim0.007$ mag for the 0.7-m and slightly lower for the 1.1-m telescope.

\section{Analysis}\label{sec:analysis}

\subsection{Disk Properties from ALMA}

The 1.3 mm continuum image in Figure \ref{fig:alma} clearly shows circumstellar disks associated with each stellar component. We fit the continuum map for each component using \texttt{imfit} in \texttt{CASA}, which provides the peak and integrated fluxes presented in Table \ref{tab:params}. 
The total integrated flux (from both disks, $4.4\pm0.26$ mJy) is consistent with lower angular resolution observations for the blended system \citep[$3.4\pm1.7 {\rm mJy}$, ][]{Andrews2013}. Assuming the dust emission is optically thin, we can compute a dust mass: 
\begin{equation}
    M_{\rm dust} = \frac{F_\nu d^2}{\kappa_\nu B_\nu(T_{\rm dust})}
\end{equation}
where $F_\nu$ is the integrated $230$ GHz flux, $d$ is the distance to the source, $\kappa_\nu$ is the dust opacity, \citep[$2.3 \cm^2\g^{-1}$ at 230 GHz,][]{Andrews2013} and $B_\nu(T_{\rm dust})$ is the Planck function evaluated at the dust temperature, $T_{\rm dust}$, which is computed as $T_{\rm dust} \sim 25 (L/L_\odot)^{1/4} \K$ \citep[also assuming optically thin dust emission]{Andrews2013}. This yields an estimate of the dust mass of $4.1\pm0.3\times10^{-6} M_\odot$ and $3.5\pm0.3\times10^{-6} M_\odot$ for the primary and secondary, respectively. These are lower limit estimates of the overall dust mass contained in the disk because the inner disks are likely optically thick \citep{Huang2018,Dullemond2018} and may contain up to an order of magnitude more dust hidden below the optical surface \citep{Zhu2019}. 

\begin{figure*}
    \centering
    \includegraphics[width=0.95\textwidth]{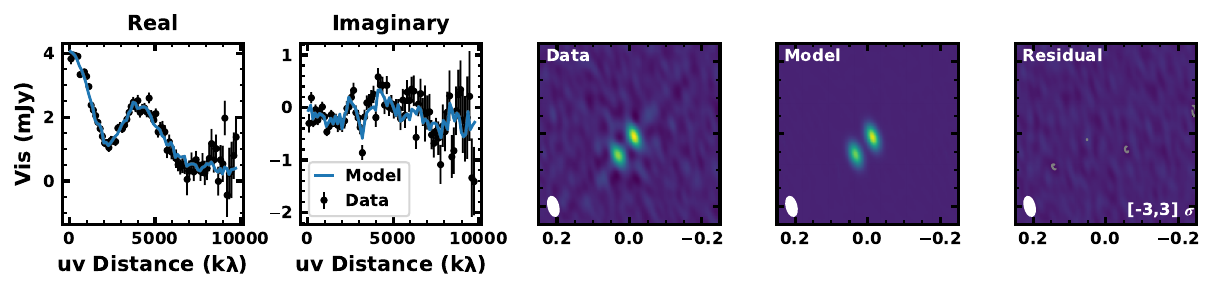}
    \caption{Continuum visibility fitting. Results are shown for spectral window 1. {\bf Left Panels:} Data and model for the real and imaginary visibilities as a function of uv-distance. {\bf Right Panels:} Maps of the data, model, and residuals. Contours in the residual image are set at $-3\sigma$ and $3\sigma$ in dashed and solid lines, respectively. The residuals are largely $<\pm3\sigma$. }
    \label{fig:UVfit}
\end{figure*}

To measure the disk parameters (namely inclination, position angle, and effective radius), we forward model the continuum visibilities with an exponentially tapered power-law intensity profile,
\begin{equation}\label{eq:intensity}
    I(r) = I_0 \left(\frac{r}{R_c}\right)^{-\gamma_1} e^{-(r/R_c)^{\gamma_2}}.    
\end{equation}
Here, $\gamma_1$ is power-law index and the disk intensity drops exponentially as $e^{-r^{\gamma_2}}$ ($\gamma_2$ is the exponential-taper index) outside of the cutoff radius, $R_c$. 
This model describes disks with sharply decreasing outer profiles, characteristic of disks in close binaries \citep{Manara2019,Tofflemire2024} and compact disks \citep{Longetal2019}. The on-sky model image modifies this radial intensity profile with an inclination, position angle, and positional offset. We then compute the complex visibilities for the model at the observed $uv$ baselines using the {\tt galario} package \citep{Tazzarietal2018} and fit them to each of the three high bandwidth spectral windows independently in an MCMC framework using {\tt emcee} \citep{emcee}. The high resolution CO spectral window is excluded from our fit given its lower bandwidth and SNR.

Our fit employs 400 walkers to fit the 12 parameter model: $I_0$, $R_c$, $i$, $PA$, and positions for each disk. The remaining radial profile indices are fixed at values characteristic of truncated binary disks \citep[$\gamma_1 = 0.8$ and $\gamma_2=5$; based on estimates in][]{Tofflemire2024,Manara2019}. The convergence of the fit is measured from the chain auto-correlation time. The first 5 auto correlation times are removed as burn in. 

We fit each of the three continuum spectral windows separately and combined the results in order to capture uncertainties beyond the instrumental measurement precision \citep[e.g.,][]{Longetal2019}. Individual fit posteriors are broad and generally overlap. Conservatively, we  combined the last 5000 steps from each spectral window fit and take the median and 95\% confidence interval as our adopted value and uncertainty. For DF Tau A, the fit returns $i_A = $\incA\incAerr$^\circ$,\ $PA_A =$ \PAA\PAAerr$^\circ$, and $R_{c,A} = $\RcA\RcAerr AU. For DF Tau B, we find $i_B = $\incB\incBerr$^\circ$, $PA_B =$ \PAB\PABerr$^\circ$, and $R_{c,B} = $\RcB\RcBerr AU. Since our exponential cutoff is sharp (because $\gamma_2$ is large), the radius containing 95\% of the total flux is $\simeq R_c$.
Figure \ref{fig:UVfit} presents a representative fit for one of the continuum spectral windows. We did not detect any dust emission that would indicate the presence of a circumbinary disk.






The $^{12}$CO maps show a weak detection around the primary and the secondary as well. 
Overall, the inferred position angle from the CO Maps (approx 120-155$^\circ$) is consistent with that from our visibility modeling. 
The $^{12}$CO emission in the north-west portion of the secondary (i.e., the would-be red-shifted emission from the companion) is absent from both the intensity and first moment maps. This is likely due to molecular cloud absorption, and is consistent with the velocity of the Taurus molecular cloud.  

\begin{deluxetable}{lll}
\tabletypesize{\scriptsize}
\tablecaption{Derived Properties for DF Tau \label{tab:params}} 
\tablehead{ \colhead{Parameter} &\colhead{DF Tau A} & \colhead{DF Tau B}}
\startdata 
\multicolumn{3}{c}{Orbital Parameters}\\
\hline
$P$ [yr]        & \multicolumn{2}{c}{$48.1 \pm 2.1$}     \\
$T_0$ [JY]      & \multicolumn{2}{c}{$1977.7  \pm 2.7$}  \\
$e$             & \multicolumn{2}{c}{$0.196 \pm 0.024$}    \\
$a$ [mas]       & \multicolumn{2}{c}{$97.0  \pm 3.2$}      \\
$i$ [$^\circ$]  & \multicolumn{2}{c}{$54.3 \pm 2.4$}      \\
$\Omega$ [$^\circ$]  & \multicolumn{2}{c}{$38.4  \pm 2.5$}      \\
$\omega$ [$^\circ$]  & \multicolumn{2}{c}{$310.6 \pm 9.2$}      \\
$M_{\rm tot}~(\frac{d}{D})^3 ~M_\odot$ & \multicolumn{2}{c}{$1.15 \pm 0.03 \pm 0.48$\textsuperscript{a}}  
\vspace{2pt}\\ 
\hline
\multicolumn{3}{c}{Stellar Properties}\\
\hline
$T_{\rm eff}$ [K]            & $ 3638  \pm 109 $ & $ 3433  \pm 84  $ \\
$\log g$                     & $  3.7  \pm 0.2 $ & $  3.9  \pm 0.2 $ \\
$v\sin i [\km/\s]$                    & $ 16.4  \pm 2.1 $ & $ 46.2  \pm 2.8 $ \\
Veiling at $15600 {\rm\AA} $ & $  1.4  \pm 0.2 $ & $  0.3  \pm 0.2 $ \\
$RV [\km/\s]$                & $ 19.9  \pm 1.1 $ & $ 15.5  \pm 2.0 $ \\
$B [{\rm kG}]$                     & $  2.5  \pm 0.7 $ & $  2.6  \pm 0.9 $  \\
JD & \multicolumn{2}{c}{2455171.91488}
\vspace{2pt}\\
\hline
\multicolumn{3}{c}{Disk Properties}\\
\hline
1.3mm $F$ [mJy]                    & $2.5 \pm 0.19 $ & $ 1.9 \pm 0.18 $\\
1.3mm Peak $I$ [mJy/beam]          & $1.5 \pm 0.08 $ & $ 1.3 \pm 0.08 $\\
$M_{\rm dust} [M_\oplus] $         & $1.4\pm0.10$    & $1.17\pm1.11$ \\
$i_{\rm disk}[^\circ]$             & \incA\incAerr & \incB\incBerr \\
${\rm PA}_{\rm disk}[^\circ]$              & \PAA\PAAerr   & \PAB\PABerr \\ 
$R_{\rm c} [\AU]$           & \RcA\RcAerr   & \RcB\RcBerr  
\vspace{2pt}\\
\hline\hline
\enddata 
\tablenotetext{a}{Based on a distance estimate of 142.68 pc to the D4-North subgroup in Taurus \citep{Krolikowski2021}. The first uncertainty in $M_{\rm tot}$ is propagated from the uncertainties in the orbital parameters $P$ and $a$ while the second systematic uncertainty is derived from propagating the $\pm$ 20 pc standard deviation of distances to individual stars in the subgroup.}
\end{deluxetable} 

\subsection{Orbital Properties from AO Imaging}
\begin{figure}
    \centering
    \includegraphics[width=0.45\textwidth]{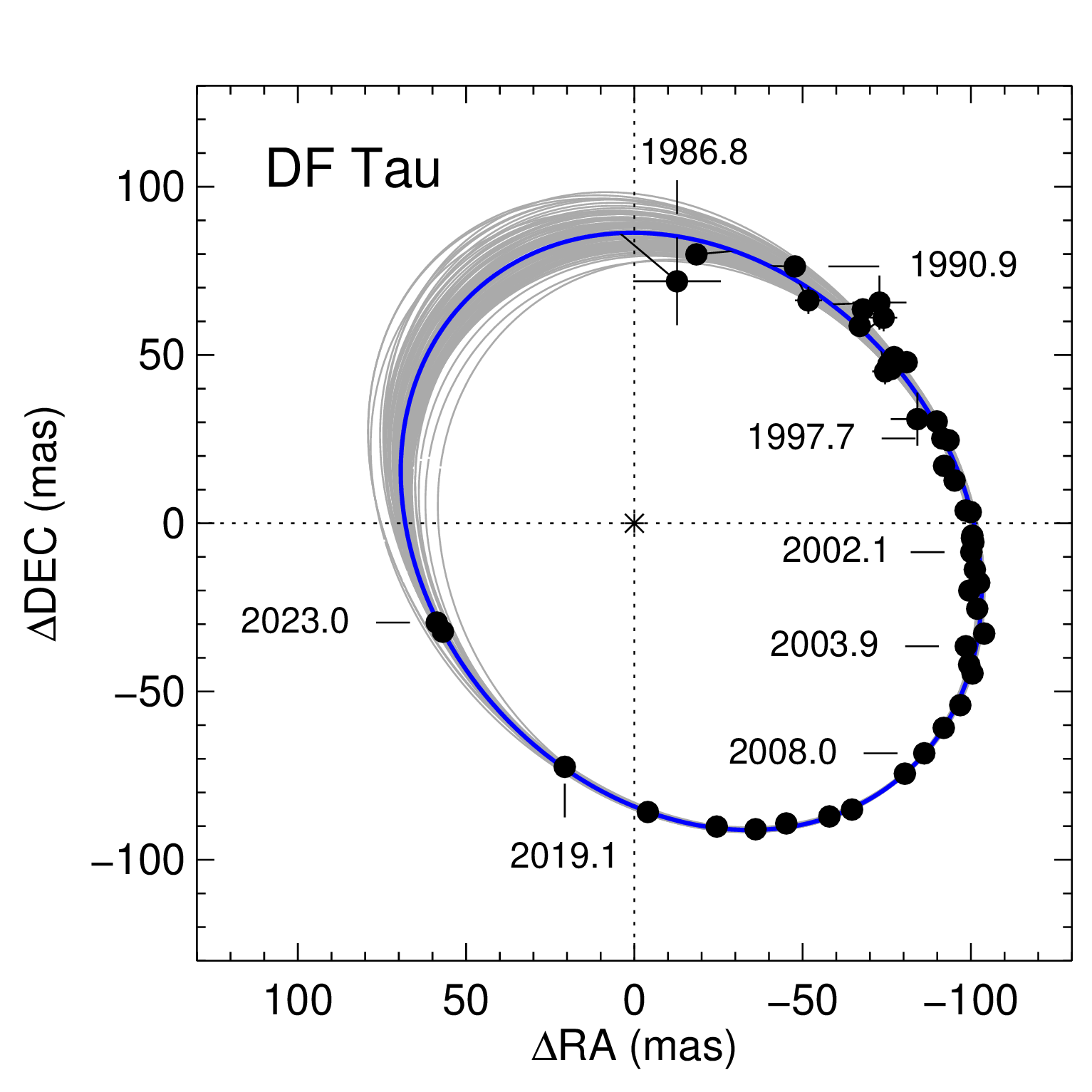}
    \caption{Orbital motion of DF Tau B relative to A based on the AO imaging and measurements from the literature. The solid blue line shows the best fitting orbit while the grey lines show 100 orbits selected at random from the posterior distribution. }
    \label{fig:orbit}
\end{figure}

The orbital coverage of DF Tau now spans over 36 years. We computed an updated orbital fit by combining the binary positions in Table~\ref{tab.sepPA} with measurements previously published in the literature \citep{Chen1990,Ghez1995,Simon1996,Thiebaut1995,White2001,Balega2002,Balega2004,Balega2007,Shakhovskoj2006,Schaefer2003,Schaefer2006,Schaefer2014,Allen2017}. We used the IDL orbit fitting library\footnote{\url{http://www.chara.gsu.edu/analysis-software/orbfit-lib}} to compute a visual orbit using the Newton-Raphson method to linearize the equations of orbital motion and minimize the $\chi^2$. The period ($P$), time of periastron passage ($T_0$), eccentricity ($e$), angular semimajor axis ($a$), inclination ($i$), position angle of the line of nodes ($\Omega$), and the angle between the node and periastron ($\omega$) are fitted for and results are presented in Table~\ref{tab:params}. The uncertainties in the orbital parameters were computed from a Monte Carlo bootstrap technique. This process involved randomly selecting position measurements from the sample with repetition (some measurements were repeated, others were left out), adding Gaussian uncertainties to the selected sample, and re-fitting orbit. We performed 1,000 bootstrap iterations and computed uncertainties in the orbital parameters from the standard deviation of the resulting distributions. The orbital motion of DF Tau B relative to A and the best fitting orbit are plotted in Figure \ref{fig:orbit}.

The total system mass can be computed from Kepler’s Third Law, assuming that the distance is known. For example, the Gaia distance changed from 124.5 pc in DR2 to 182.4 pc in DR3 with a Renormalized Unit Weight Error (RUWE) of 21.9 \citep{Bailer-Jones2018,Bailer-Jones2021}. The high RUWE is likely due to the binarity of the source and a more reliable distance will be obtained when the final solution that includes the astrometric orbital motion is available. However, the variability of DF Tau A \citep{Allen2017} might impact the interpretation of the photo-center motion.  In the meantime, we adopted the distance of $142.68\pm20$ pc to the D4-North subgroup where DF Tau resides  \citep{Krolikowski2021} to derive a total mass of $M_{\rm A + B} = 1.15 \pm 0.48 M_\odot$, where uncertainties propagated from the orbital parameters $P$ and $a$ ($\pm 0.03 M_\odot$) and the distance ($\pm 0.48 M_\odot$) are added in quadrature.  


\subsection{Stellar Properties from NIR Component Spectroscopy} 

We used the NextGen atmospheric models \citep{Allard1995} with the Synthmag spectral synthesis code of \citet{Kochukhov2010} to produce a grid of H-band order 49 ($1.5440-1.5675 \, \mu$m) model spectra at solar metallicity that encompasses a large range of late-type dwarf star properties: $T_{\rm eff}$ of $3000–6000$ K, surface-averaged magnetic field strength, B, of 0–6 kG, and surface gravity, $\log (g)$, of $3.0-5.5$. Models are computed at intervals of $100 \K$ in $T_{\rm eff}$ and $0.5$ dex in $ \log (g)$ for magnetic field strengths of 0, 2, 4, and 6 kG.
Laboratory atomic transition data from the Vienna Atomic Line Database 3 \citep[VALD 3, ][]{Ryabchikova2015} were calibrated against the spectrum of 61 Cyg B and the Solar spectrum \citep{Livingston1991} and used to synthesize the model spectra following the procedure of \citet{Johns-Krull1999,Johns-Krull2007}. 
We excluded lines which appeared in the solar and 61 Cyg B calibration spectra but not in the VALD list, and vice versa. 
Furthermore, we adjusted the van der Waals broadening constants and line oscillator strengths in the model spectra to best match the observed calibrator lines in the Sun and 61 Cyg B to improve the accuracy of the synthesized spectra.

From this grid of model spectra, we can linearly interpolate to create model spectra with any desired stellar parameters spanned by the grid. We determined the best-fit stellar parameters using the \texttt{emcee} implementation of the MCMC method. To identify an appropriate initial guess, we used the stellar parameters estimated by \citet{Prato2023b}. We also applied the line equivalent width ratio method of \citet{Tang2024} to obtain initial temperature estimates of $\sim$3690 K and $\sim$3630 K for the primary and secondary, respectively. However, we suspect this is a temperature overestimate for the secondary, given the line blending that results from this star's high $v\sin i$. Finally, we adopt a wide range of uniform priors for all stellar parameters. 

\begin{figure}
    \centering
    \includegraphics[width=0.45\textwidth]{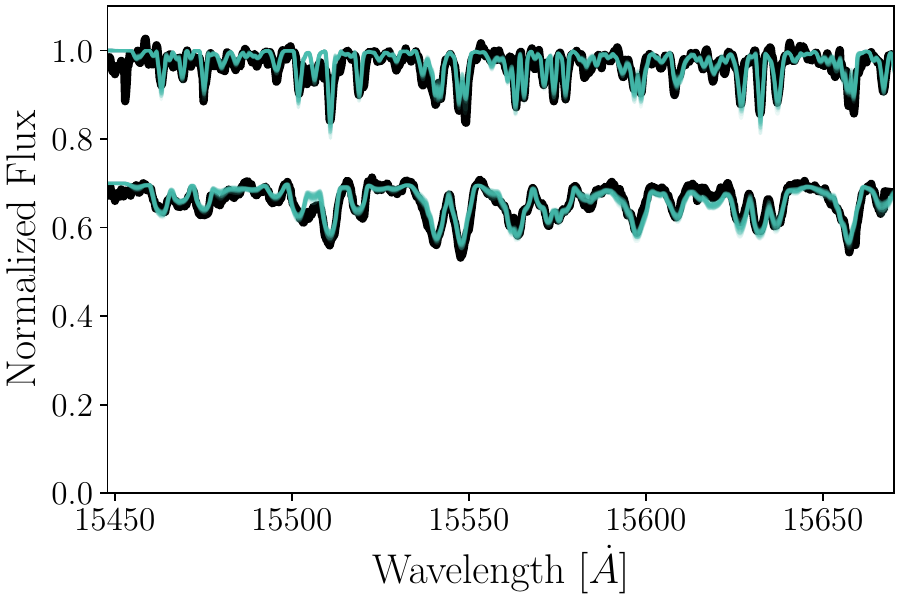}
    \caption{Spatially resolved and continuum normalized spectra of DF Tau A (top) and B (bottom, separated by an offset) in black. 50 randomly selected model spectra from the trimmed MCMC chain are overplotted in blue.}
    \label{fig:specfit}
\end{figure}

With our set of initial guesses and priors, we ran \texttt{emcee} with $32$ walkers and iterated for 8000 steps, which allows for a chain length that is sufficiently longer than the autocorrelation time. We then trimmed the first $2000$ steps before extracting model parameters. Our best-fit models are shown in Figure \ref{fig:specfit} and the associated model parameters are listed in Table \ref{tab:params}. 

The primary and secondary of DF Tau are almost stellar twins, with similar effective temperatures, surface gravities, and surface-averaged magnetic field strengths. The large differences for the two stars lie in their veiling and $v\sin i$ values. The lack of veiling led \citet{Allen2017} to conclude that either the disk around the secondary had dissipated or the inner disk was absent. This is supported by the lack of accretion signatures in Figure \ref{fig:specfit}, e.g., Br 16 H line emission. Examination of other H-band spectral orders shows no trace of other Brackett series lines (e.g., Br 11 in order 45 or Br 13 in order 47).

Given our independent measurements for stellar effective temperatures from the spectra and a combined dynamical mass from the orbit, we can also compare mass estimates from evolutionary models. Using the \citep{Feiden2016} stellar evolution models, the stellar masses which correspond to stellar effective temperatures of $3638$ K and $3433$ K are $0.56 M_\odot$ and $0.42 M_\odot$, respectively. This is consistent with our estimate measured total mass of the system, $1.15\pm0.48 M_\oplus$. We can also compare the model-derived flux ratio to that determined in Table \ref{tab.sepPA}. \citet{Feiden2016} models give logarithmic luminosities of $-0.33$ and $-0.48$, for the primary and secondary respectively, which corresponds to a flux ratio of 0.69, in rough agreement with the ratios in Table \ref{tab.sepPA}. 

\subsection{Stellar Variability and Rotation Properties from Time Series Photometry}

Peaks with significant power in periodograms are used to determine stellar rotation periods, leveraging the light curve modulation caused by spots rotating in and out of view \citep[e.g., ][]{Rebull2020}. For DF Tau, both the space-based and ground-based lightcurves appear to be dominated by stochastic variability (Figure \ref{fig:periodograms}). The periodogram of the TESS lightcurve reveals two significant peaks at $10.4$ and $5.5$ days. The periodogram of the $K2$ lightcurve has peaks at $10.3$ and $6.6$ days; however there is significantly more power at $15.2$ and $21.8$ days. The periodogram of six seasons' worth of ground-based photometry has low power peaks at approximately $8$ and $14$ days. 

\citet{Allen2017} analyzed one season of ground-based photometry and found a peak in the periodogram at $10.4$ days, but they did not find significant power at periods of $5.5$ or greater than $10.4$ days. It is not clear that any of these light curves are reliably tracking the stellar rotation period, given the implied stellar radius. 

For example, when a suspected rotation period, identified in the periodograms derived from TESS, $K2$, or ground-based light curves, is combined with a $v\sin i$, determined from the primary star H-band spectrum (e.g., Section \ref{subsec:keck-nirc2}), we can obtain a lower bound on the stellar radius: 
\vspace{-5pt}
\begin{equation}\label{eq:rstar}
    R_* \geq 0.0196 \Bigg(\frac{P_{\rm rot}}{1 \,{\rm day}}\Bigg)\Bigg(\frac{v\sin i}{1{\rm \,km/s}}\Bigg) R_\odot . 
\end{equation}
\vspace{1pt}\\ 

For a measured $v\sin i$ of $13 \pm 4 {\rm km/s}$ and $P_{\rm rot}=10.4$ days, \citet{Allen2017} found a stellar radius limit for the primary star of $\geq 2.68 \pm 0.82 R_\odot$. Using our updated $v \sin i$ of $16.4  \pm 2.1 \km/\s$, a 10.4 day period implies a radius of $\geq 3.4 \pm 0.4 R_\odot$, implying an age well below 1 Myr for the models of \citet{Feiden2016}. For the secondary star with $v\sin i$ of $46.2 \pm 2.8 {\rm km/s}$, the derived radii are unphysically large for rotation periods greater than $\sim$3 days.

To estimate these limits on the radii above, Equation \ref{eq:rstar} assumed a maximum stellar rotation axis inclination of 90 degrees. Another estimate of the stellar radii can be achieved by assuming the star is aligned with the disks or the orbit of the binary (i.e., an inclination of $i=34-55^\circ$). For DF Tau A, with assumed age of $\sim$2 Myr and $T_{\rm eff} \sim 3650 \K$, models of \citet[][which include stellar magntic fields]{Feiden2016} predict a stellar radius of $\sim 1.8 R_\odot$. With $v \sin i$ of $16.4  \pm 2.1 \km/\s$, these parameters imply a rotation period range from 3.1 to 4.5 days for the primary star. 
For DF Tau B, with assumed age of $\sim$2 Myr and $T_{\rm eff} \sim 3450 \K$, models predict a stellar radius of $\sim 1.6 R_\odot$. With $v \sin i$ of $46.2 \pm 2.8 \km/\s$, these parameters imply a rotation period range from 1.0 to 1.4 days for the secondary star for the inclination range of $i=34-55^\circ$.
We find no significant power at predicted periods of $<$5 days in the periodogram for any of the light curves shown in Figure 3. Stellar evolution tracks which do not include stellar magnetic fields, such as \citet{Baraffe2015} or \citet{Feiden2016}'s non-magnetic tracks, have smaller radii on average, which further broadens these discrepancies.  

\begin{figure}
   \centering
   \includegraphics[width=0.45\textwidth]{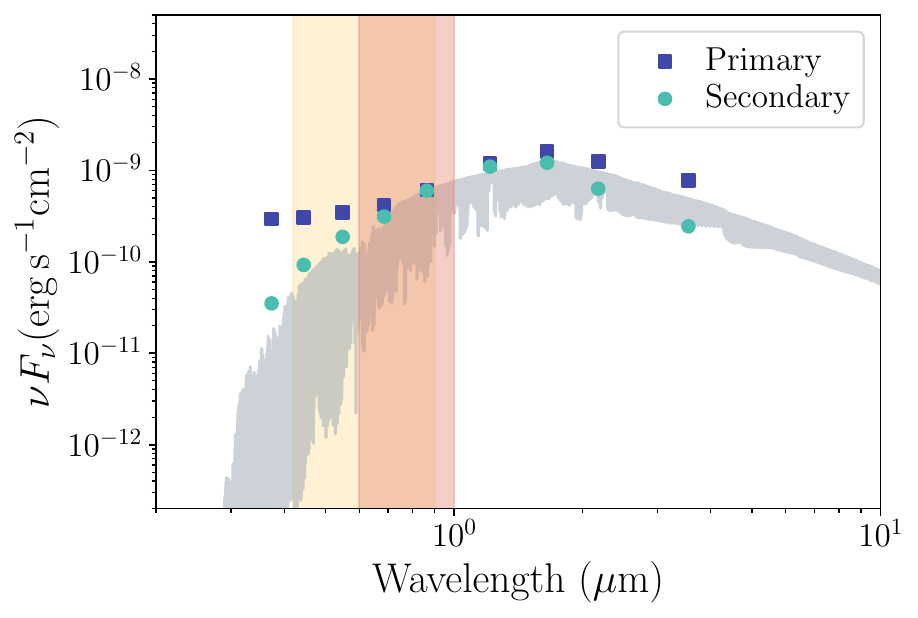}
   \caption{Angularly resolved photometry of DF Tau A (squares) shows a clear UV Excess when compared with DF Tau B (circles) and an appropriate BT-Settl (CIFIST) stellar photosphere model (grey). Yellow and orange regions denote the Kepler and TESS bandpasses, respectively. Photometry is reproduced from \citet{Allen2017}, error bars are smaller than the markers. }
   \label{fig:SED}
\end{figure}

It is therefore unlikely that the periodic signals seen in the TESS, $K2$ and ground-based light curves consistently trace stellar rotation. Instead, it is more probable that these lightcurves are dominated by stochastic accretion events onto the primary star. 
In the TESS  and $K2$ bandpasses, shown in Figure \ref{fig:SED}, the flux ratios of the primary to the secondary are $\sim1.2$ and $\sim1.9$. This suggests that the light curves are indeed dominated by the actively-accreting primary. 
This conclusion is also supported by seasonal variation of the lightcurve, as seen in our ground-based data presented in Appendix \ref{ap:season}.

\section{Discussion}\label{sec:discuss}
We discuss our measurements of the components of DF Tau and their protoplanetary disks in the context of both the binary-disk interaction as well as possible origins for the missing inner disk of the secondary. 

\subsection{Binary-Disk and Disk-Disk Interactions}

As in \citet{Tofflemire2024}, our finding of (rough) agreement in the projected inclinations ($41\pm10^\circ$ and $46\pm9^\circ$ for the primary and secondary, respectively) and position angles (\PAA\PAAerr$^\circ$ and 
\PAB\PABerr$^\circ$) of the circumstellar disks and the orbit of this young binary ($i=54.3\pm1.2^\circ$, $\Omega = 38.4\pm2.5^\circ$) does not distinguish between formation mechanisms. Quantatively, we compute the disk-orbit obliquity ($\Theta$) for this system: 
\begin{equation}\label{eq:obliq}
\begin{split}
    cos\Theta & = \cos i_{\rm disk}\cos i_{\rm orbit} \\ 
    & + \sin i_{\rm disk} \sin i_{\rm orbit} \cos(\Omega_{\rm disk} - \Omega{\rm orbit}),
\end{split}
\end{equation}
where $\Omega_{\rm disk}$ is the position angle of the disk. We find the disk-orbit obliquity to be consistent with zero for both components: the obliquities are $13\pm13^\circ$ and $8\pm9^\circ$ for the primary and secondary disk and binary orbit. We can also use Equation \ref{eq:obliq} to compute the disk-disk obliquity, which is also consistent with zero ($\Theta=5\pm16^\circ$).

Disk-orbit alignment can be achieved from either formation due to fragmentation of a gravitationally unstable disk \citep[e.g. ][]{Bate1997,Ochi2005,Young2015} or from an initially misaligned system, formed from core fragmentation \citep[e.g. ][]{Zhao2013,Lee2019,Guszejnov2023}, which has undergone significant damping from various physical mechanisms \citep[see ][ and references therein for details]{Offner2023}. 
The damping mechanism that is relevant on the scales of the binary orbit of DF Tau is viscous warped disk torques \citep[e.g.][]{Bate2000,Lubow2000}. The damping timescales for these torques are short and decrease $\propto a^6 r_{\rm out}^{9/2}$ \citep{Zanazzi2018}. The small disk radii, $r_{\rm out}$, and semi-major axis of the binary orbit, $a$, in the DF Tau system makes a damping timescale smaller than the age of the system \citep[$\sim 2$ Myr from][]{Krolikowski2021}. 

Given the tight orbital separation of DF Tau, another meaningful comparison is with theoretical predictions of the truncation radius \citep{AL1994,Lubow2015,Miranda2015}, beyond the first order approximation of $R_{\rm trunc}\simeq a/3$. Following \citet{AL1994}, \citet{Manara2019} derive the truncation radius as: 
\begin{equation}
    R_{\rm trun}=\frac{0.49\,a\,q^{-2/3}}{0.6q^{-2/3} + \ln(1+q^{-1/3})} \times \big(be^c+0.88\mu^{0.01}\big) \,, 
\end{equation} where $a$ is the semi-major axis, $q$ is the stellar mass ratio ($M_B/M_A$), $e$ is the orbital eccentricity, and $\mu$ is the secondary to total mass ratio ($M_B/(M_{\rm A+B})$). The parameters $b$ and $c$ depend on $\mu$ and the disk Reynolds number. Assuming the stars are of equal mass, and using $b=-0.78$ to $-0.82$ and $c=0.66$ to $0.94$, the ranges in Appendix C1 of \citet{Manara2019}, we obtain truncation radii estimates of $3.1$--$3.6 \AU$ for DF Tau's orbital parameters. 

Dust disk radii that are smaller than this dynamical prediction are likely the result of dust drift \citep{Ansdell2018}. 
The effects of dust drift on disks in binaries are further enhanced because the outer reservoir of dust particles is lost during disk truncation \citep{Zagaria2021,Rota2022}. 
Compared to the radii we measure of 3.7 and 3.6 AU (i.e. the 95\% effective radius in Table \ref{tab:params}) for the primary and secondary respectively, there is little evidence of radial drift. However, since our disks are only marginally resolved it is possible that the disks are slightly misaligned from the binary orbit, allowing for larger gaseous extents. But since CO maps are of poor quality, we cannot assess whether the gas disk sizes are consistent with theoretical predictions. Higher angular resolution continuum observations or higher SNR CO maps would make this measurement more robust. 

\subsection{Stellar Obliquity and Rotation}
The large $v\sin i$ of the secondary is also consistent with our picture of DF Tau B lacking an inner disk. The effects of disk locking \citep{Shu1994}, which slows the rotation of stars with strong magnetic fields frozen in to the inner disks, have weakened or dispersed for the secondary but modulate the slow rotation of the primary. 

Our best-fit for the disk inclinations ($41\pm10^\circ$ and $46\pm9^\circ$ for the primary and secondary, respectively) is consistent with alignment to the binary orbit ($54.3\pm1.2^\circ$). Given this relative co-planarity, it is not improbable that the binary orbit, the circumstellar disks, and the stellar obliquity are all aligned. If this is the case, and we assume a stellar spin axis that is aligned with the orbital angular momentum vector of the binary orbit, we can estimate a rotation period for the secondary star, assuming an age $\sim$2 Myr, of $\lesssim 1.4$ days. For this orientation, we also estimate that the velocity at the surface of the secondary is only $\sim 20 \%$ of the breakup velocity. 

Comparing to the rotation periods of disk-hosting and disk-less single stars in Taurus obtained by \citep{Rebull2020}, this estimate is on the shorter side of the distribution but it is not implausibly fast. Furthermore, there are significantly more stars with rotation periods $<2$ days that do not have disks than those that do \citep[e.g. those in Figure 8 of ][]{Rebull2020}. This is most likely because the stars have spun up after the loss of their disks. Finding DF Tau B in this region of parameter space that bridges the gap between disked and disk-less systems is sensible considering that it appears to be in the early stages of disk dissipation. 


\subsection{Circumbinary Disk}

Given the accretion rate of the primary star, the lack of a circumbinary disk in this system is also puzzling. \citet{Hartigan2003} measure an accretion rate to be $\sim 10^{-7} M_\odot/{\rm year}$.  First, we assume this accretion is entirely due to the primary because the secondary does not show a UV excess. For a dust to gas ratio of 1\%, and assuming a constant accretion rate, the circumstellar disk around the primary should disappear in $\sim 3000$ years. This dissipation timescale is orders of magnitude smaller than the current age of the system($\sim 2$ Myr) and therefore in order for the circumstellar disk around the primary star to remain, another mass reservoir is necessary. This problem is not resolved if the circumstellar disk is significantly optically thick, which only adds a factor of up to 10 to the estimated lifetime. Can this timescale be extended significantly if the circumstellar disks are supplemented by an outer circumbinary disk? This question prompts us to determine how much dust could be hidden below our detection limit in a circumbinary disk.

\begin{figure}
   \centering
   \includegraphics[width=0.45\textwidth]{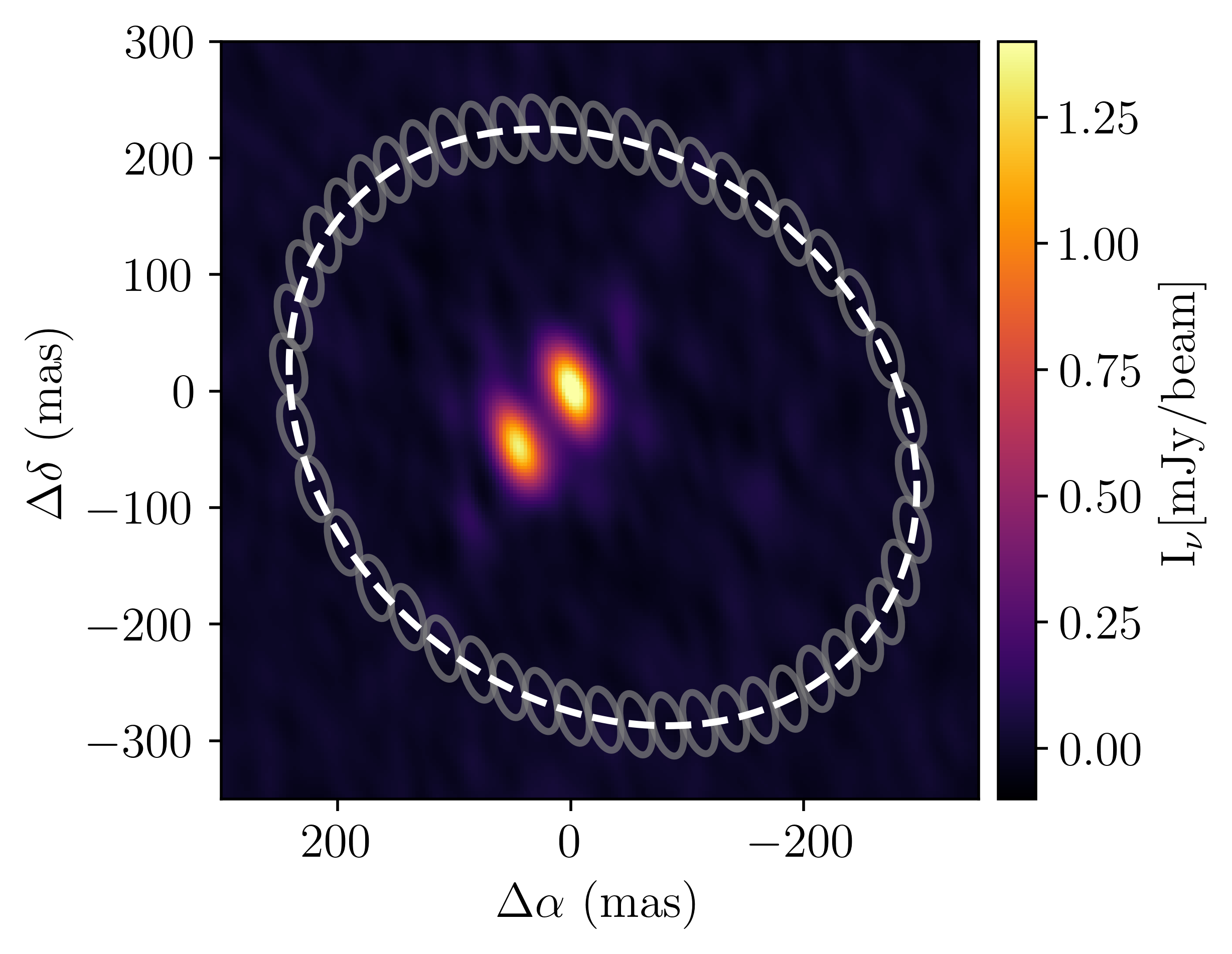}
   \caption{The expected location of the inner edge of a potential circumbinary disk (white dashed line) lies at approximately 3 times the semi-major axis of the binary orbit and is centered on the approximate center of mass of the binary. This does not show significant emission in our final continuum map. However, we place an upper limit on the amount of dust hidden in the circumbinary disk by tracing its inner edge with 48 beams (overplotted grey ovals).  }
   \label{fig:cb}
\end{figure}

In order to place an upper limit on the amount of dust contained in the inner region of a circumbinary disk, we first calculated the predicted radius of the inner edge of the circumbinary disk. This can be estimated using the orbital parameters and a semi-major axis 3 times that of the binary orbit. Figure \ref{fig:cb} shows the expected location of the circumbinary disk. We estimated the flux around the circumbinary disk in two ways. First, we arranged ellipses which have sizes and orientations the same as the beam in order to determine that approximately 50 beams are required to cover the inner edge of the circumbinary disk. Using the RMS of the continuum image of $0.013$ mJy/beam, we estimate the noise in this region as RMS$\sqrt{N_{\rm beams}} = 0.09$ mJy.  

To check this result, we summed the integrated flux for each of the beams in Figure \ref{fig:cb} to get an estimate of the integrated flux for a donut-shaped region encompassing the potential circumbinary disk. We found the integrated flux in this region to be $0.27$ mJy. When combined with our integrated fluxes for the two disks, this estimate is still consistent with previous measurements of the unresolved system\citep{Andrews2013}. This flux corresponds to a dust mass of $1.5\times 10^{-7} M_{\odot}$ ($0.15 M_\oplus$) for the circumbinary disk. If all this dust fed only the primary disk, the expected lifetime would still be only hundreds of years. 

We only measured the dust emission in a narrow ring near the expected inner edge of the binary disk because that is where the dust will be most concentrated. With an estimated projected separation for the inner disk edge of 0.3” and a maximum recoverable scale of  0.67", our observations have the capability to detect a circumbinary disk at this distance if it is present. However, the lack of a detection suggests that the dust may be in a more diffuse and extended disk which our observations are less sensitive to. Therefore the dust estimates above are likely underestimates of the total dust mass contained in the circumbinary disk. 

Now, we return to our lifetime problem for the circumstellar disk. If all the material in the circumbinary disk is directed to only the circumstellar disk of the primary, the expected lifetime would only increase by an additional $\sim450$ years. Therefore it is unlikely that the circumbinary disk is a significant additional mass reservoir for the circumstellar disks.

\subsection{Twin Stars, Mismatched Disks}
Central to the puzzle of DF Tau is the question of why the inner disk of the secondary has dissipated when the stars should be coeval.
One possibility is that these disks are at different stages of dissipation. Here, we entertain possible origins for such physical differences, in terms of the initial mass of the disk and the viscosity, and discuss whether making the necessary measurements is feasible. 

The initial mass of circumstellar disks is roughly proportional to the stellar mass \citep{Andrews2013} with a large variance. Perhaps if the disk around the primary initially held significantly more mass, equal accretion and thus dissipation rates might have produced the mismatched disks. Although our estimates of the disk dust mass are roughly the same for both outer disks, optically thick emission can hide a significant amount additional mass and the true masses of the disks may be significantly different \citep{Zhu2019}. Other methods to determine the masses of the disks face similar difficulty. For example, kinematic determinations of disk mass, which assume the self gravity of the disk is significant \citep[e.g. ][]{Veronesi2021}, may allow for more accurate estimates. Unfortunately these estimates are only accurate for disks with masses $>5\%$ of the stellar mass, and require high spatial resolution molecular maps \citep[e.g.,][demonstrate that \>20 beams across the disk are needed for an accurate estimate]{Andrews2024}. Given the truncation of both disks in DF Tau, the disks are not likely to be a substantial fraction of the stellar mass and the disks are too small to achieve the necessary spatial resolution ($\sim$0.002") for even the most extended ALMA configuration (0.01"). 
Therefore it is unlikely this method can give an insightful determination of the true disk mass. 

Another method by which we might determine the origin of the uneven dissipation between the disks in DF Tau is by determining the disks' viscosity, as parameterized by the dimensionless Shakura-Sunyaev $\alpha$ parameter \citep{ShakuraSunyaev1973}. During early disk dissipation, the accretion rate is determined by the gas viscosity: increased viscosities correspond to more accretion, leading to a depleted inner disk. However, determining whether the viscosity of the two disks is significantly different is not an easy task. 

Direct determinations of $\alpha$ require spatially and spectrally resolved line emission \citep[e.g.,][]{Flaherty2015,Teague2016} in order to disentangle thermal and turbulent line broadening. For the dim and compact disks in DF Tau, observations of this sort will have prohibitively long integration times and therefore are unlikely to be performed. Common model-dependent measurements of $\alpha$ require either resolved annular substructures \citep[e.g.,][]{Dullemond2018} or constraints on the dust's vertical scale height \citep[e.g.][]{Villenave2020,Doi2021}. Because our current observations do not reveal evidence for annular substructures in the image or visibility planes, and our disks are not edge on, these methods also will not allow for a determination of the disk viscosity. However, interferometric observations that resolve the inner cavity of DF Tau B, which may be possible with the most extended ALMA configurations, could allow for an indirect measurement of the disk viscosity using methods similar to \citet{Dullemond2018}. 

Another possibility which could clear out the inner disk of the secondary is an unknown (sub-)stellar companion. Although the presence of an undetected companion is possible, it may be unlikely in this system. Super-Earth-like planets may be able to carve gaps in low viscosity disks \citep{Dong2017,Dong2018}, but the solid core mass required to form a super-Earth is substantial ($\sim5-10 M_\oplus$). Although the measured dust masses of the circumstellar disks are small (1.4 and 1.2 $M_\oplus$ for the primary and secondary, respectively), perhaps a disk which is optically thick, and thus contains substantially more dust which is hidden, could supply the material required to build these cores. If a planet has already formed in the circumstellar disk of the secondary, perhaps that is the cause of the difference in observed dust masses between the two disks. A planet with a mass of only 0.2$M_\oplus$ likely will not have a large impact on the shape of the circumstellar disk. However if the disk were optically thick, then perhaps the missing mass in the secondary, compared to the primary, can be more similar to that of a super-Earth. 

Because there may be enough solid material in the secondary’s small circumstellar disk to form a planet capable of altering the shape of an inner disk, we do not rule out this possibility. However,  we also do not favor this interpretation because of the need for a significant amount of additional dust mass in the disk for previous or ongoing planet formation. Planets are rare in the closest binary systems \citep{Kraus2016} and a planet embedded in the circumstellar disk around DF Tau B would certainly make this system uncommon. However, clearer and more direct evidence is certainly needed to substantiate such a proposal for this disk.

Although it is challenging to determine why the evolution of DF Tau's disks is not the same, it is possible to continue to characterize the physical differences of the disks. For example, the $3.5 \mu$m magnitude places a lower limit on the possible inner edge of the secondary: dust grains that have $\lambda_{\rm peak}=3.5\mu$m must be heated to $\sim800 \K$. Since these temperatures will occur in the disk midplane interior to $1 \AU$ \citep[$T(r=1 \AU) = 150\K$ for a passively heated disk][]{Chiang1997}, the dominant heat source is accretion \citep{DAlessio1998}. Therefore we determine the radius of dust which would dominate emission at this wavelength, and constrain the inner edge of the secondary's disk to be $\gtrsim0.25 \AU$. 
Spatially resolved imaging of the secondary at $10\mu{\rm m}$ will better constrain the inner edge of its disk. Furthermore, the peculiarities of DF Tau make it an excellent candidate to test theories of disk dissipation. Angularly-resolved spectroscopic observations of wind-sensitive lines, like the He\,\textsc{i} $1083$ nm line, provide a sensitive probe of stellar and inner disk winds, which can be compared to those predicted from theory \citep[e.g.,][]{Pascucci2023}. 

In connection with our results, we note that another study of DF Tau by \citet{Grant2024}; their preprint appeared while we were completing this manuscript. Independently, \citet{Grant2024} used data from our ALMA program to
identify the disk around the secondary and came to the conclusion that the inner disk of DF Tau B is absent.

\section{Conclusions \& Summary }\label{sec:summ}
We analyzed the orbital, stellar, and protoplanetary disk properties of the puzzling young binary system, DF Tau, using decades of high angular resolution imaging, angularly-resolved NIR spectra and (sub-)mm interferometry. With the expectation of co-evolution for the disks in young binary systems, we discuss how DF Tau may be a prime target to begin to test disk dissipation theory. The main conclusions of our work are as follows: 
\begin{enumerate}
    \item DF Tau is a young binary system with a semi-major axis of $\sim 14 \AU$. Its components are twin stars with a dynamically-determined total mass of $\sim 1.1 M_\odot$ and similar effective temperatures. 
    \item Our ALMA observations in Band 6 detect continuum and \CO $J=2-1$ line emission from each component. The disks have maximum radii $4 \AU$, the result of tidal truncation. The integrated $1.3 {\rm mm}$ fluxes are $2.5\pm0.19$ and $1.9\pm0.18$ mJy, which allow for a lower estimate of the dust masses of $1.4\pm0.10 M_\oplus$ and $1.2\pm0.11 M_\oplus$ for the primary and secondary, respectively. These low dust masses offer little support to any ongoing planet formation.  
    \item The inclinations and position angles of the dust disks, measured by directly fitting the visibilities, point to mutual alignment between the binary orbit and the circumstellar disks. 
    \item We do not find evidence for a circumbinary disk or extended emission in dust or CO above an RMS of $0.013$ and $1$ mJy/beam, respectively. We constrain the upper limit of the dust emission from the circumbinary disk to be $<0.27$ mJy.  
\end{enumerate}

The accelerated dissipation of the inner, terrestrial planet-forming zone in the DF Tau B circumstellar disk has important implications for planet formation. This system merits intensive future scrutiny in order to better understand the conditions that disrupt, in the case of DF Tau B, and sustain, for DF Tau A, the circumstellar disks over a timescale relevant for planet formation.

\section{Acknowledgements}
The authors thank Feng Long, Mike Simon, J.J. Zanazzi, Peter Knowlton, Meghan Speckert, and Jacob Hyden for helpful discussions. We are also grateful to Joe Llama and Cho Robie, who generously shared their computational resources.

Funding for this research was provided in part by NSF awards AST-1313399 and AST-2109179 to L. Prato. G. Schaefer and L. Prato were also supported by NASA Keck PI Data Awards, administered by the NASA Exoplanet Science Institute. 

Some of the data presented herein were obtained at the W. M. Keck Observatory from telescope time allocated to the National Aeronautics and Space Administration through the agency's scientific partnership with the California Institute of Technology and the University of California. Additional telescope time was awarded by NOIRLab (NOAO and NOIRLab PropIDs: 2009B-0040 and 2022B-970020; PI: G. Schaefer) through NSF’s Telescope System Instrumentation and Mid-Scale Innovations Programs. The Keck Observatory is a private 501(c)3 non-profit organization operated as a scientific partnership among the California Institute of Technology, the University of California, and the National Aeronautics and Space Administration. The Observatory was made possible by the generous financial support of the W. M. Keck Foundation. The authors wish to recognize and acknowledge the very significant cultural role and reverence that the summit of Maunakea has always had within the indigenous Hawaiian community. We are most fortunate to have the opportunity to conduct observations from this mountain.

This paper makes use of the following ALMA data: ADS/JAO.ALMA\#2019.1.01739.S ALMA is a partnership of ESO (representing its member states), NSF (USA) and NINS (Japan), together with NRC (Canada), MOST and ASIAA (Taiwan), and KASI (Republic of Korea), in cooperation with the Republic of Chile. The Joint ALMA Observatory is operated by ESO, AUI/NRAO and NAOJ. The National Radio Astronomy Observatory is a facility of the National Science Foundation operated under cooperative agreement by Associated Universities, Inc.

\facility{ALMA, Texas Advanced Computing Center (TACC), Keck:II (NIRSPEC)}
\software{
\texttt{emcee} \citep{emcee}, \texttt{lightkurve} \citep{lightkurve}, \texttt{astropy} \citep{astropy:2013,astropy:2018,astropy:2022}, \texttt{galario}\citep{Tazzarietal2018}, \texttt{casa} \citep{casa}
}
\bibliography{sample631}{}
\bibliographystyle{aasjournal}

\appendix
\section{Seasonal Variation of the Periodogram} \label{ap:season}
\begin{figure}
    \centering
    \includegraphics[width=0.85\textwidth]{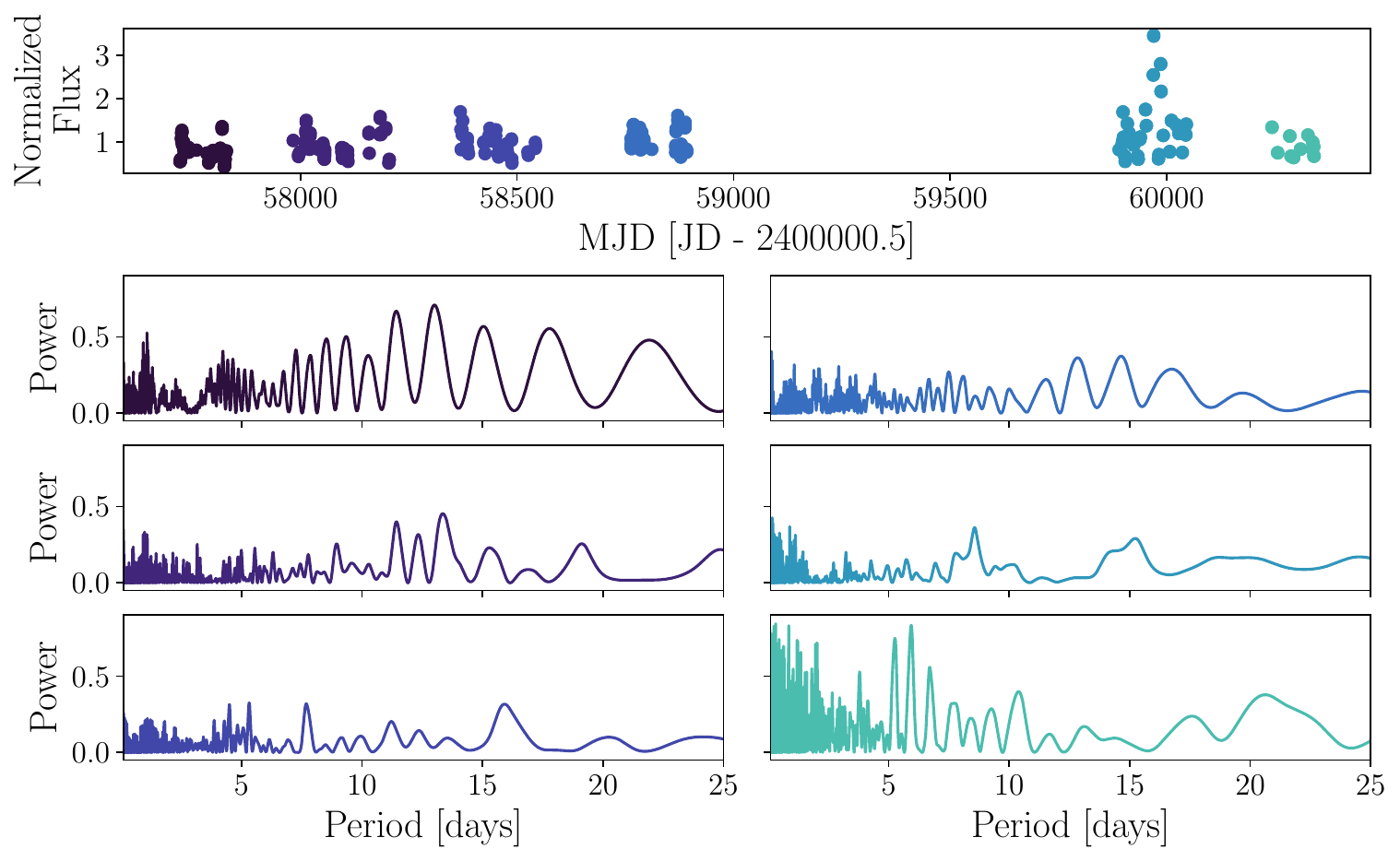}
    \caption{Seasonal variation of the DF Tau periodogram using ground-based photometry obtained with the Lowell robotic 0.7m telescope and the Lowell Hall 1.1m telescope.}
    \label{fig:season}
\end{figure}

To support our claim that the variation in the light curves of DF Tau is dominated by stochastic accretion, we analyzed the 7-year-long ground-based light curve of DF Tau separated by year (Figure \ref{fig:season}). The strength of the $\sim10$- and $\sim5$-day signals varies, and at some epochs disappears entirely (e.g., in the periodograms from the seasons around $58000$ and $60000$ MJD). 
\label{app:season}

\end{document}